\documentclass[prl,aps,twocolumn,superscriptaddress]{revtex4}
\usepackage{graphicx, amsmath, amsthm, subfigure}
\makeatletter
\begin{document}
\title{Exploring Exotic Superfluidity of Polarized Ultracold Fermions in Optical Lattices}

\author{Yan Chen}
\affiliation{Department of Physics and Center of Theoretical and
Computational Physics, The University of Hong Kong, Pokfulam Road,
Hong Kong, China} \affiliation{Department of Physics and Lab of
Advanced Materials, Fudan University, Shanghai, China}
\author{Z. D. Wang}
\affiliation{Department of Physics and Center of Theoretical and
Computational Physics, The University of Hong Kong, Pokfulam Road,
Hong Kong, China} \affiliation{National Laboratory of Solid State
Microstructures, Nanjing University, Nanjing, China}
\author{F. C. Zhang}
\affiliation{Department of Physics and Center of Theoretical and
Computational Physics, The University of Hong Kong, Pokfulam Road,
Hong Kong, China} 
\author{C. S. Ting}
\affiliation{Department of Physics and Texas Center for
Superconductivity, University of Houston, Houston, TX 77204, USA}

\date{\today}






\maketitle


\textbf{Enormous interest has been paid to ultracold Fermi gases due
to the interplay between Cooper pairing and strong
correlations~\cite{Fermi1,Fermi2,Fermi3}. Beautiful experiments on
the superfluidity have been performed in these systems with unequal
spin
populations~\cite{Partridge1,Partridge2,Zwierlein1,Zwierlein2,Zwierlein3,Zwierlein4}.
Arrestingly, it was found that the superfluid paired core is
surrounded by a shell of normal unpaired fermions while the density
distribution of the difference of the two components becomes
bimodal. Here we explore theoretically the novel superfluidity of
harmonically-trapped polarized ultracold fermionic atoms in a
two-dimensional (2D) optical lattice by solving the Bogoliubov-de
Gennes equations.
The pairing amplitude is found to oscillate along the radial
direction at low particle density and along the angular direction at
high density. The former is consistent with the existing experiments
and the latter is a newly predicted Fulde-Ferrell-Larkin-Ovchinnikov
(FFLO) state, which can be tested in experiments.}

Experimentally, optical lattices artificially created by
standing-wave laser fields~\cite{Lattice1,Lattice2,Lattice3} provide
unprecedented experimental tunability, such as  a precise control of
model Hamiltonian parameters including the interaction strength,
lattice geometry and number of atoms. This versatile tool opens a
door to explore rich and novel physical phenomena in various
strongly correlated systems. Experimental exploration of the novel
superfluid phase of imbalanced fermions in optical lattices is
expected to be conducted in the near future, while in the
theoretical aspect, there exists a long-standing debate on the
nature of the exotic pairing state for the imbalanced systems. The
inhomogeneous superconducting state, known as the FFLO state, was
predicted about forty years ago, which contains nonzero momentum
Cooper pairs with the superconducting gap exhibiting a periodic
real-space modulations~\cite{FFLO}. There have been suggestions for
the possible FFLO phase for some of the high-$T_c$, layered organic,
and heavy fermion superconductors at very low temperatures and high
magnetic fields ~\cite{FFLO-expts1,FFLO-expts2}. This subject has
also been studied in nuclear and high-energy
physics~\cite{Inhomo-SC}. Despite of many previous literatures in
this
field~\cite{WLiu,Machida06,Sheehy,Torma06,Strinati,Bulgac,Melo,Stoof,Mueller,Yip,LMDuan,QJChen},
it is still awaited to explore theoretically and unambiguously the
exotic FFLO state with imbalanced ultracold fermions in optical
lattices.

In this letter, we present an unrestricted Hartree-Fock study of a
two-component fermionic atoms in a harmonically confined 2D
optical lattice. Due to the competition between
the ferromagnetic and $s$-wave superfluid phases in the presence
of harmonically trapped potential, the compromised ground state
may exhibit rich patterns of order parameter configurations. Our
main findings are as follows. (i) In the low fermionic filling
regime, the superfluid-pairing gap modulates its sign along the
radial direction, indicating an FFLO state. The calculated
density profiles of the two components of the fermions show a striking resemblance to
the experimental observations. (ii) In the high filling regime,
a superfluid-pairing ring appears.  As  the imbalance increases,
the pairing gap oscillates along the angular direction, signaling
a novel FFLO state. (iii)  A variety of intriguing distributions
of the order parameter,
 such as square lattice FFLO state, emerges in the intermediate filling regime.
In addition, we also observe  that, as the imbalance increases,
the superfluidity is suppressed and vanishes at a critical
imbalance amplitude, implying the existence of the Clogston
limit~\cite{Clogston}. 

To simulate the neutral fermionic atoms in a 2D optical lattice,
we begin with an effective tight-binding Hamiltonian, which
captures the basic interplay between ferromagnetism and $s$-wave
fermion-pairing in the presence of a confining potential,
\begin{eqnarray}
\hat{H}& = &-t\sum_{\langle
ij\rangle,\sigma}\hat{c}_{i\sigma}^\dagger \hat{c}_{j\sigma}
+\sum_{i\sigma} (\epsilon_{i}-\mu_{\sigma}) \hat{n}_{i\sigma}\\
& + & \sum_{\langle ij \rangle} \left[ \Delta_{i}
\hat{c}^\dagger_{i\uparrow} \hat{c}^\dagger_{i\downarrow} +
\mbox{h.c} \right].\nonumber
\end{eqnarray}
where $\hat{c}_{i\sigma}$ is an annihilation operator for an atom at
site $i$ (at position ${\mathbf{r}}_i$) with spin $\sigma$ and
$\hat{n}_{i\sigma}=\hat{c}_{i\sigma}^\dagger \hat{c}_{i\sigma}$ is
the number operator.  $t$ is the effective nearest-neighboring
hopping integral. The spin-dependent chemical potential due to the
Zeeman splitting is defined as $\mu_{\sigma}= \mu - \sigma h$.
$\epsilon_{i}=\frac{1}{2}m\omega^2 |\mathbf{r_i}-\mathbf{r_0}|^2$ is
the harmonic confining potential at site $i$,  and $\mathbf{r_0}$ is
the position of the trap center.

At the mean-field level, the model Hamiltonian can
be diagonalized by solving the resulting Bogoliubov-de Gennes (BdG)
equations self-consistently~\cite{BdG1,BdG2},
\begin{equation}\label{BdG}
\sum_j \left( \begin{array}{cc} H_{ij,\sigma} & \Delta_{ij}^*  \\
\Delta_{ij} & -H_{ij,\bar\sigma}^*
\end{array} \right) \left( \begin{array}{c} u_{j\sigma}^n \\
v_{j\bar\sigma}^n
\end{array}
\right) = E_{n} \left( \begin{array}{c} u_{i\sigma}^n \\
v_{i\bar\sigma}^n \end{array} \right).
\end{equation}
where $H_{ij,\sigma} =
-t_{ij}-(\epsilon_i+ \mu_{\sigma})\delta_{ii}$
is the single particle Hamiltonian, $\Delta_{ij} =
\Delta_{i} \delta_{ii}$ is the pairing order parameter,
and $(u_{i\sigma}^n,v_{i\bar\sigma}^{n})$
are the Bogoliubov quasiparticle amplitudes at the site $i$.
The atom density and the $s$-wave pairing order parameter
satisfy the self-consistency conditions, $n_{i \uparrow} =
\sum_{n} |u_{i \uparrow}^n|^2 f(E_n)$, $n_{i\downarrow} = \sum_{n}
|v_{i \downarrow}^n|^2 [1-f(E_n)]$ and $\Delta_{i} = V
\sum_{n}(u_{i\uparrow}^nv_{i\downarrow}^{n*} +
u_{i\uparrow}^nv_{i\downarrow}^{n*})\tanh(E_n/2k_BT)$,
respectively, where $T$ is the temperature.
The local magnetization is defined as
$m_i=n_{i\uparrow}-n_{i\downarrow}$, which can be tuned by
changing the magnetic field in the present model.
In a laboratory optical lattice,
$m_i$ can be tuned directly although there is no real magnetic field.
In our calculations below, we set $t=1$ as the energy unit and lattice spacing $a=1$
as the  length unit. We consider a very low temperature $T=0.001$,
and the amplitude of confining potential
$\frac{1}{2}m\omega^2 a^2/t \sim (0.016 - 0.025)$.
The dimension of lattice is $42 \times 42 $ with open
boundary conditions and the trap center is located at $(21,21)$.
All the order parameters or mean fields are calculated self-consistently by
solving the BdG equations iteratively, with their initial values
randomly distributed on the lattice. In the case of multi-solutions,
we compare their corresponding free energies to determine the most
energetically favored state.

The applied magnetic field may cause the density imbalance of the
two species of fermions (spin-up and spin-down). The induced
ferromagnetism competes with the superfluid pairing state, resulting
in the emergence of exotic FFLO phases. In the special case
$\epsilon_i=0$  and in 2D, the pairing gap of the ground state of
Hamiltonian (1) oscillates in space with a stripe-like pattern for
the s-wave and a square lattice pattern for the
$d$-wave~\cite{CSTing}. In the crossing region of the positive and
negative pairing amplitudes, there exist nodal lines where the
ferromagnetism appears. The presence of the trapped potential may
add up to the new complication. In the widely used local density
approximation (LDA), the spatial dependence of observable must
follow contours of constant trapping potential in the same way that
a spatially uniform system does. Our approach provides a consistent
way to explore novel states beyond LDA. In particular, our results
show that the compromising ground state  depends crucially on the
$n_c$ (fermion density at the trap center). The features presented
below can be classified into three regimes accordingly.

\begin{figure}
\centering
\includegraphics[width=\columnwidth,height=6.5cm]{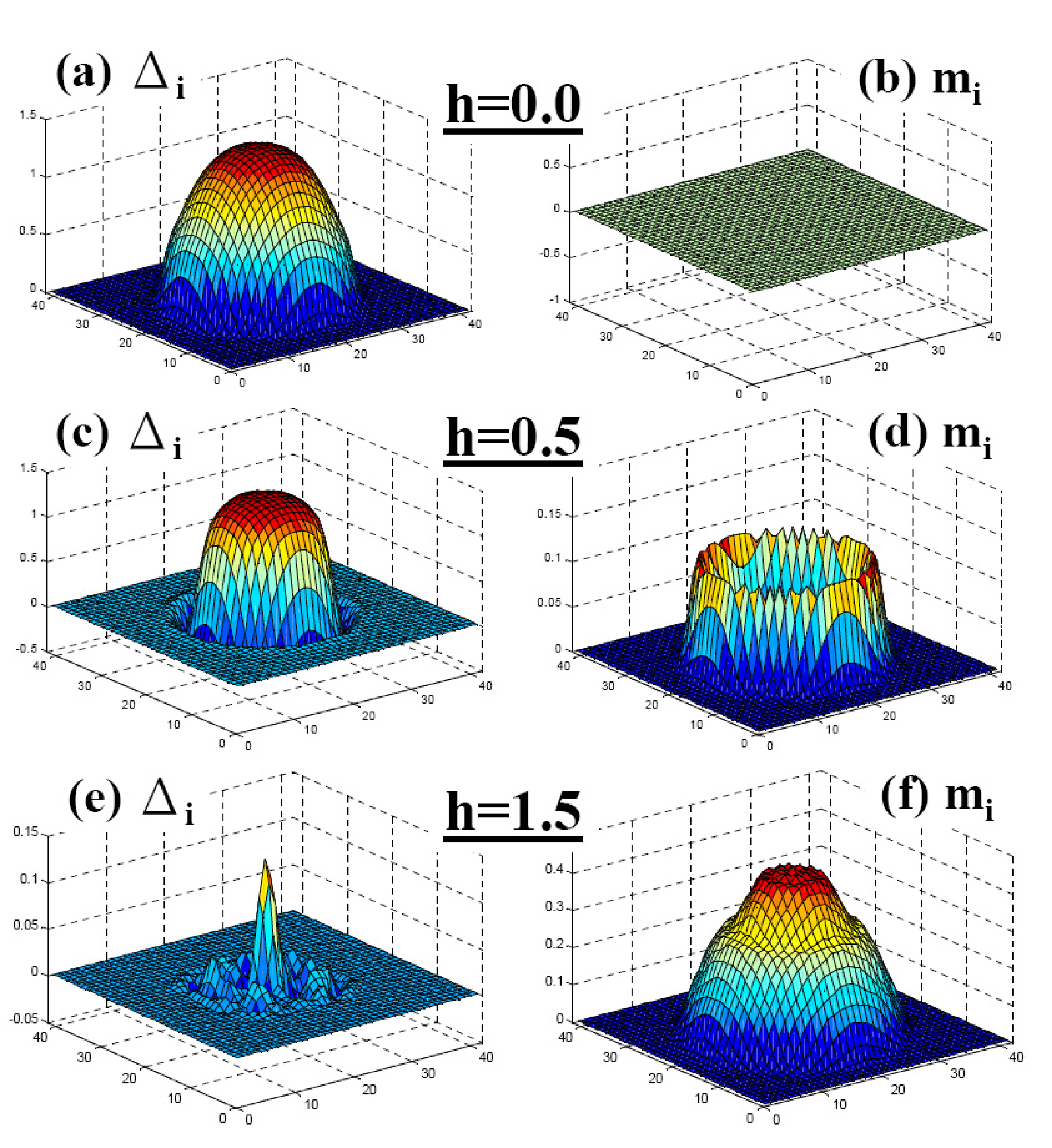}
\caption{(Color online) Spatial profiles of pairing order parameter
$\Delta_i$ [(a), (c), and (e)], and local magnetization $m_i$ [(b),
(d), and (f)] of Eqn. (1) at a low particle filling.  $N=200$
fermions on a lattice of $42 \times 42$. The parameter is
$\frac{1}{2}m\omega^2 a^2/t =0.016$. The magnetic field is (top to
bottom): $h=0, 0.5, 1.5$, respectively. } \label{figure1}
\end{figure}
\emph{Low-filling regime ($n_c <1$)}. In Fig. 1, we plot typically
three spatial profiles of the pairing order parameter $\Delta_i$
together with local magnetization $m_i$ as functions of magnetic
field $h$. At $h=0$, there is no ferromagnetic order
and the trapping potential may result in the accumulation of
electron density around the center.  As shown in Fig. 1(a)-1(b),
$\Delta_i$ reaches its maximum value at the trap center and
decreases to zero far away from the center. These results are
consistent with those obtained by LDA, where no competing orders
show up. As $h$ increases, the ferromagnetic order may
emerge and frustrate the conventional pairing state. In Fig. 1(c), our
calculations show the appearance of superfluid pairing gap
modulations along the radial direction, indicating an FFLO state.
At the edge of the modulation, a continuous sign change of
the pairing gap corresponds to a nodal line ($\Delta_i=0$).
Meanwhile the magnetic order parameter $m_i$ is most
remarkable around the ring-like nodal line so that its density
profile exhibits a bimodal structure, as shown in Fig. 1(d). In
other words, the minority fermions are squeezed into the inner
core due to the pair condensation in the superfluid phase, while the
unpaired majority fermions are repelled to the outside of the
core. When the imbalance increases, the minority fermions contribute
more effectively to the pairing states. This result is in good
agreement with the experiments, and also with the previous theoretical
studies on the FFLO state in a continuous model~\cite{Machida06,Torma06}.
As $h$ progressively increases, the pairing order may
further be suppressed and the weak modulation of the pairing  gap
appears away from the trap center while the ferromagnetic order
shows a plateau-like feature nearby the trap center.
Furthermore, the superfluidity vanishes at a critical
magnetic field ($h_c \sim 1.6$), which corresponds to the Clogston
limit. Note that the existing experimental realization
~\cite{Partridge1,Partridge2,Zwierlein1,Zwierlein2,Zwierlein3,Zwierlein4}
may correspond to the present low-filling case. The
density profiles obtained in our theory exhibit remarkable resemblance to the
experimental measurements.

\begin{figure}
\centering
\includegraphics[width=3.2in,height=4.5cm]{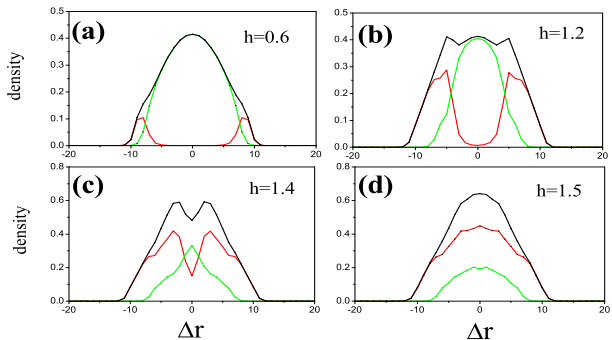}
\caption{(Color online) Density profiles along the diagonal of the lattice.
$\Delta r$ is the distance from the trap center.
Red line denotes the magnetization while the black and
green lines represent the spin-up and spin-down densities,
respectively. All the parameters are the same as in Fig. 1} \label{figure2}
\end{figure}
To show the \emph{bimodal} structure more transparently, we display
the density profiles along the lattice diagonal through the trap
center. The density distribution has a four-fold symmetry,
consistent with the underlying lattice structure. As shown in Fig.
2, the bimodal distribution of magnetization clearly shows up at
finite $h$. There are three regions: a superfluid core with equal
densities, a partially polarized shell, and a fully polarized
region. By increasing $h$, these bimodal structures evolve to have
more pronounced amplitude while the separation between peaks becomes
narrower. As $h$ exceeds a critical value, the equal density core
disappears and  the condensation fraction also vanishes. In such a
case, the ferromagnetic order reaches its maximum at the trap center
and its density profile displays a plateau-like structure.

\begin{figure}[t]
\includegraphics[clip=true,width=\columnwidth,height=7.8cm]{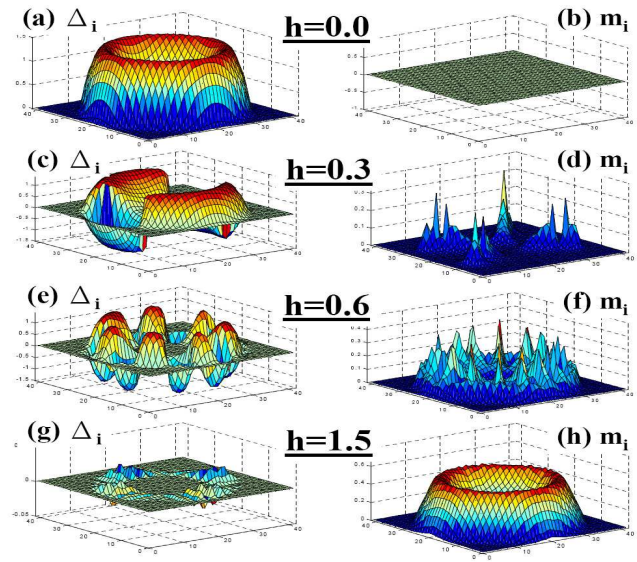}
\caption{(Color online) Spatial variations of $\Delta_i$ [(a), (c),
(e) and (g)], and local ferromagnetic order $m_i$  [(b), (d), (f)
and (h)] at a high filling: $N=1200$ fermions. Parameters are the
same as in Fig. 1.
} \label{figure3}
\end{figure}

\emph{High-filling regime ($n_c$ close to 2)}. In this regime, the
band is almost fully occupied, corresponding to an \emph{insulating}
state in a conventional solid state system, around the trap center.
The pairing order can not survive far way from the center because of
the low fermion density. In the intermediate distance from the
center, one may naturally expect the formation of ring like fermion
pairing state at the intermediate range. In Fig. 3(a)-3(b), the
profile of $\Delta_i$ does show such a topological structure at
$h=0$. Upon increasing $h$, the ferromagnetic order may emerge and
result in the frustration of the competing pairing state. It is well
known that the lower angular momentum state has always the lower
energy in the infinite system. However, in the present case, due to
the highly nontrivial interplays among trapping potential,
magnetization and pairing correlations, it is possible that the
exotic pairing state with higher angular momentum state becomes
energetically favorable. As depicted in Fig. 3(c), the pairing gap
oscillates along the angular (ring) direction with alternating
positive and negative
signs. 
In Fig. 3(d), it shows clearly four regions with abrupt sign change
where magnetization shows up. This special configuration lowers the
free energy and allows oscillating fermion pairing state to remain
stable. By further increasing $h$, there will be more oscillations
along the angular direction as well as more appreciable
magnetization, as displayed in Fig. 3(e)-3(f). When $h=1.5$, the
superconducting order is almost fully suppressed while the
ferromagnetic order forms a shell pattern [Fig. 3(g)-3(h)].


An intuitive physical understanding of angular FFLO state can be
given as follows: the main effect of the high filling at the trap
center is the formation of superconducting shell structure as
shown in Fig. 3(a). As is known, the FFLO is rather robust in
1D~\cite{1D1,1D2}. If the superconducting shell is sufficiently
thin like a 1D ring, then FFLO tends to emerge. That is to say,
the spin imbalance will lead to the pairing oscillation along the
angular direction rather than in the radial direction since the
former state costs less kinetic energy. This novel angular FFLO
state can be observed in 2D optical lattices where high filling
regime can be easily achieved. We have checked our results for the
$d$-wave pairing interaction, and the similar conclusions have
also been reached. It is worth to mention that the simple LDA
fails to get such a novel pairing state. Here we need to stress
the critical role played by trapping potential in the formation of
the FFLO states along radial or angular direction. The appearance
of the FFLO state minimizes the relevant free energy, which is a
direct manifestation of the competition between ferromagnetism and
superconductivity. The trapping potential provides a confined and
inhomogeneous background, which complicates the competition, but
reveals some of the novel physics associated with the FFLO states.

\begin{figure}[t]
\includegraphics[clip=true,width=\columnwidth,height=9.0cm]{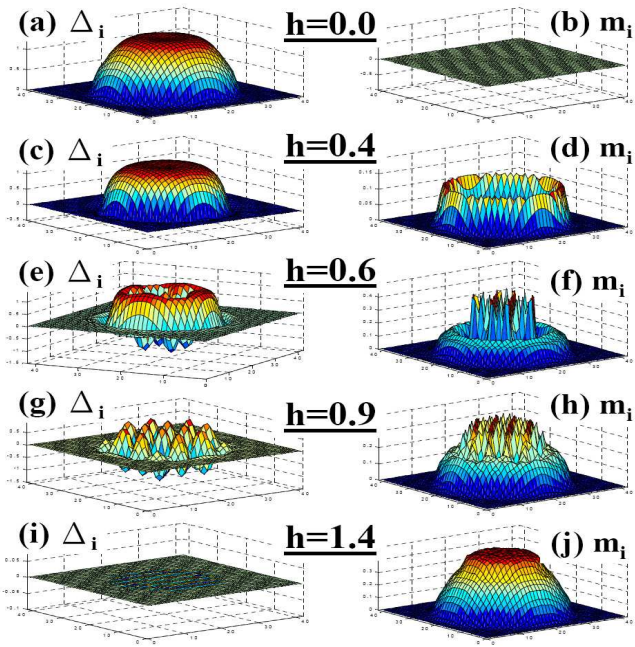}
\caption{(Color online)  Spatial profile of (a) pairing order
parameter, and (b) magnetization $m_i$ at an intermediate filling:
$N=400$ fermions. The parameter is $\frac{1}{2}m\omega^2 a^2/t
=0.025$ and other parameters are the same as in Fig. 1.
} \label{figure4}
\end{figure}
\emph{Medium-filling regime ($n_c \in (1,2)$).} In this regime, rich
patterns of order parameter distribution show up.  In the absence of
the magnetic field, the pairing order parameter shows a small
concaved structure around the trap center, as seen in Fig. 4(a). By
tuning the magnetic field to $h=0.4$, the pairing gap modulation
along the radial direction shows up and the situation is analogous
to the low filling case [Fig. 1(d)-1(f)]. When $h=0.6$, the pairing
order parameter modulates not only along the radial direction but
also around the trap center, exhibiting a square lattice pattern as
illustrated in Fig. 4(e). Meanwhile, its ferromagnetic order
exhibits a rich structure where there are many local maxima around
the trap center which correspond to nodal lines in Fig. 4(f). As we
further increases the magnetic field to $h=0.9$, the square lattice
pattern of FFLO state becomes more pronounced and the associated
ferromagnetism exhibits itself as well [Fig. 4(g)-4(h)]. Note that
the square lattice FFLO state appears in the infinite 2D system with
the $d$-wave pairing~\cite{CSTing}. When $h$ becomes quite large,
the BCS pairing order is fully suppressed and the magnetization
displays a plateau like structure around the trap center, as
depicted in Fig. 4(i)-4(j).

So far, there has been no direct evidence of FFLO states in recent
experiments since the probe of pairing gap distribution is very
challenging. We now address an important issue on how to detect the
peculiar real space patterns illustrated in Figs. 3 and  4
 experimentally. First we propose a 2D optical
lattice experiment with the fermionic atoms in the high filling
regime. By increasing the imbalance progressively, we expect the
emergence of angular dependent magnetization distribution,
indicating the existence of angular FFLO state. Very recently, the
MIT group produced preliminary experimental evidence for
superfluidity of ultracold $^6$Li atoms in optical
lattice~\cite{Lattice3}. We expect that a future experiment for
imbalanced fermions will be conducted to examine our prediction. An
alternative scenario to detect such an exotic state is to carry out
the experiment in a thin superconducting ring of heavy fermion
materials. By applying strong magnetic field parallel to the plane,
the angular-dependent FFLO state may show up due to the special
topology of the ring structure. Probe of the real space modulation
of the pairing gap as well as magnetization can be achieved by using
SQUID or STM techniques. The third proposal is to generate a Mexican
hat trapping potential in 3D so that the distribution of the
confined fermionic atoms may form a donut-like structure. In such a
case, the angular dependent distribution of magnetization profile
can be directly measured by using the time-of-flight imaging
techniques.

In summary, we have explored theoretically the novel superfluidity
and ferromagnetism in ultracold fermionic atoms on a 2D lattice
combined with a harmonic trap.
Due to the interplay between the confined ferromagnetic and
superfluid orders, exotic FFLO phases have been revealed.
At low density, our theory shows a
bimodal distribution of the magnetization profile and a
fermion pairing gap oscillating along the radial direction,
reproducing the main features observed in
recent experiments. We predict a more exotic
angular FFLO state at high densities and square lattice like
FFLO state at the intermediate densities.

\vspace{0.3in} \textbf{Acknowledgements}

This work was supported by the RGC grants of Hong Kong
(HKU7012/06P, HKU7051/06P, and HKU-3/05C), Seed Funding grants of
HKU, the National Natural Science Foundation of China (10429401),
the state key programs of China (2006CB925204), and the Robert A.
Welch Foundation under grant E-1146.



\end{document}